# Iterative Adaptively Regularized LASSO-ADMM Algorithm for CFAR Estimation of Sparse Signals: IAR-LASSO-ADMM-CFAR Algorithm

Huiyue Yi, Yan Xu, Wuxiong Zhang, and Hui Xu

*Abstract*—The least-absolute shrinkage and selection operator (LASSO) is a regularization technique for estimating sparse signals of interest emerging in various applications and can be efficiently solved via the alternating direction method of multipliers (ADMM), which will be termed as LASSO-ADMM algorithm. The choice of the regularization parameter has significant impact on the performance of LASSO-ADMM algorithm, especially when the SNR (signal-to-noise ratio) is relatively low. However, the optimization for the regularization parameter in the existing LASSO-ADMM algorithms has not been solved yet. In order to optimize this regularization parameter, in this paper we propose an efficient iterative adaptively regularized LASSO-ADMM (IAR-LASSO-ADMM) algorithm by iteratively updating the regularization parameter in the LASSO-ADMM algorithm. Moreover, a method is designed to iteratively update the regularization parameter by adding an outer iteration to the LASSO-ADMM algorithm. Specifically, at each outer iteration the zero support of the estimate obtained by the inner LASSO-ADMM algorithm is utilized to estimate the noise variance, and the noise variance is utilized to update the threshold according to a pre-defined const false alarm rate (CFAR). Then, the resulting threshold is utilized to update both the non-zero support of the estimate and the regularization parameter, and proceed to the next inner iteration. In addition, a suitable stopping criterion is designed to terminate the outer iteration process to obtain the final non-zero support of the estimate of the sparse measurement signals. The resulting algorithm is termed as IAR-LASSO-ADMM-CFAR algorithm. Since the proposed IAR-LASSO-ADMM-CFAR algorithm could learn sparsity information at each iteration, it ensures a sparser and sparser solution and obtains the final sparse solution when the designed stopping criterion is satisfied. Finally, simulation results have been presented to show that the proposed IAR-LASSO-ADMM-CFAR algorithm outperforms the conventional LASSO-ADMM algorithm and other existing algorithms in terms of reconstruction accuracy, and its sparsity order estimate is more accurate than the existing algorithms.

## I. INTRODUCTION

Compressive sensing (CS) [1]-[2] is an efficient signal processing technique for signal acquisition and reconstruction in a sub-Nyquist sampling sense, and has been made significant progress over the past decade. Moreover, it has found wide applications in many fields of signal processing, such as wideband spectrum sensing in cognitive radio (CR) systems [3]-[5], CS-based channel estimation [6]-[10], high dimensional financial index tracking [11]-[12], sparse reconstruction for bearings vibration signals [13], CS-based multitarget CFAR detection algorithm for FMCW Radar [14], SAR Imaging [15], and so on. Among the existing CS-based reconstruction algorithms, the iterative greedy pursuit algorithms have attracted significant attention due to their low complexity and competitive reconstruction performances. The conventional greedy algorithms include orthogonal matching pursuit (OMP) [16], compressive sampling matching pursuit (CoSaMP) [17], subspace pursuit (SP) [18], gradient projection method in [19], least square absolute shrinkage and selection operator (LASSO) algorithm [20], etc. These algorithms [16]-[20] require the sparsity order as a known prior information. For example, the greedy algorithms such as MP or OMP require the sparsity order as the prior knowledge to determine the number of iterations for achieving better detection performance, and the LASSO-based techniques [20]-[24] need to select the regularization parameter related to the sparsity order. In [22], the LASSO problem is solved via the alternating direction method of multipliers (ADMM), which will be termed as LASSO-ADMM algorithm hereafter. However, the actual signal sparsity order is often unknown in advance and can also vary over time in many practical applications such as in cognitive radio (CR) systems [4, 24], where the sparse signal may vary with time both in its nonzero support set as well as in the values of its nonzero entries. Moreover, their estimation accuracy and complexity would be significantly affected by the noise term, especially when the signal-to-noise ratio (SNR) is relatively low. Therefore, it will be of important practical value if the task of signal detection is completed in the case that the signal sparsity order is unknown.

One possible method to deal with the unknown and/or possibly varying sparsity is to estimate the sparsity order based on the measurement data [4], [25]-[28]. In [25], an eigenvalue-

This work was supported by Shanghai Fundamental Research Key Project under Grant 20JC1416504. *(Corresponding author: Huiyue Yi).*

The authors are with the Shanghai Institute of Microsystem and information technology, Chinese Academy of Science, 4/F, Xinwei Building A, 1455 Pingcheng Road, Jiading, Shanghai, People's Republic of China (e-mail: huiyue.yi@mail.sim.ac.cn; xy@mail.sim.ac.cn; wuxiong.zhang@mail.sim.ac.cn; hui.xu@mail.sim.ac.cn).

based approach using the eigenvalues of the measured signal's covariance matrix is proposed to estimate the sparsity order of the wideband spectrum, which can be subsequently used for implementing adaptive CS-based CR receivers. In [26], a trace-based method is proposed to estimate the sparsity order of wideband spectrum by utilizing the trace of measured signals' covariance matrix. In [27], theoretical methods were utilized to estimate the unknown sparsity order. In [4], a two-step adaptive compressive spectrum sensing (TS-ACSS) method was proposed to estimate the sparsity order. In this method, a coarse estimation of the sparsity order can be obtained by using a simple statistical analysis of the compressive measurements, and then an accurate estimation can be acquired by comparing multiple iterations. However, the sparsity order estimation methods mentioned above still require the maximum sparsity order to determine the sampling rate in the sparsity order estimation process, where extra high sampling rate is still required.

Instead of directly estimating the sparsity order of signals mentioned above, another method on CS with unknown sparsity order is to design certain criterions to stop the sensing process adaptively [6]-[7] and [29]-[32]. In [6], a sparsity adaptive CoSaMP with regularization was proposed for underwater channel estimation, which utilizes the regularization process to realize the secondary selection of atoms and can adaptively adjust the step length. In [7], an adaptive OMP algorithm was proposed for underwater acoustic OFDM channel estimation. In [14], a sparsity adaptive correlation maximization CFAR (SACM-CFAR) algorithm was proposed for target detection in an FMCW radar, which distinguishes the target from the clutter on the basis of the correlation between linear measurements of the radar IF signal and the sensing matrix. The sparsity adaptive matching pursuit (SAMP) algorithm proposed in [29] designed a reasonable iterative stopping condition to stop the sensing process and recover the signal with unknown sparsity order. However, it requires the user to specify a step size. Similarly, the autonomous compressive spectrum sensing (ACSS) scheme proposed in [30] can adjust the number of compressed measurements without any prior knowledge about the sparsity order. In [31], several stopping criteria were proposed for the WGSOMP (Wideband Group Simultaneous Orthogonal Matching Pursuit) algorithm, which aims to maximize the probability of detection while minimizing the probability of false alarm in unoccupied bands. In [32], an adaptively-regularized iterative reweighted least squares algorithm with sparsity bound learning is designed to efficiently recover sparse signals from measurements. Specifically, at each iteration the support of estimated signal is exploited to construct a sparsity-promoting matrix, and then an adaptive regularization is formulated.

The LASSO in [21] is a regularization technique for simultaneous estimation and variable selection, and the choice of the regularization parameter in the LASSO has major impact on the performance of LASSO. Specifically, the regularization parameter in LASSO is a penalty parameter that balances the reconstruction accuracy and the sparsity of the minimization result. In addition, the choice of the regularization parameter depends on the noise level in the original measurement signal. The LASSO continuously shrinks the coefficients toward 0 as the regularization parameter increases, and some coefficients are shrunk to exact 0 if the regularization parameter is sufficiently large. Therefore, the performance of the LASSO will be affected by both the regularization parameter and the noise term, especially in the case of low SNR. Therefore, it is required to find the most suitable value for this regularization parameter. In order to deal with this problem, some methods are proposed to adaptively change the regularization parameter in the LASSO [33]-[36]. In [33], an interior-point method was proposed for solving large-scale 1-regularized least-squares programs (LSPs) that uses the preconditioned conjugate gradients algorithm to compute the search direction. In [34], an adaptive LASSO was proposed for simultaneous estimation and variable selection, in which adaptive weights are used for penalizing different coefficients in the l1 penalty term. In [35], a recursive least-squares (RLS) weighted Lasso (RLS-LASSO) was proposed for recursive estimation and tracking of (possibly time-varying) sparse signals based on noisy sequential observations. In [36], a fast and scalable polyatomic Frank-Wolfe (P-FW) algorithm for the resolution of high-dimensional LASSO regression problems.

As can be seen from above discussions, the optimization for the regularization parameter in the existing LASSO-ADMM algorithms has not been solved yet. Therefore, it is difficult to guarantee the detection and false alarm probabilities with the traditional threshold setting algorithms when the noise power is unknown or fluctuates in real-time wideband spectrum sensing. In order to optimize this regularization parameter, in this paper we propose an efficient iterative adaptively regularized LASSO-ADMM (IAR-LASSO-ADMM) algorithm by iteratively updating the regularization parameter in the LASSO-ADMM algorithm. Moreover, a method is designed to iteratively update the regularization parameter by adding an outer iteration to the LASSO-ADMM algorithm. Specifically, at each outer iteration the zero support of the estimate obtained by the inner LASSO-ADMM algorithm is utilized to estimate the noise variance, and the noise variance is utilized to update the threshold according to a pre-defined const false alarm rate (CFAR). Then, the resulting threshold is utilized to update both the non-zero support of the estimate and the regularization parameter, and proceed to the next inner iteration. In addition, a suitable stopping criterion is designed to terminate the outer iteration process to obtain the final non-zero support of the estimate of the sparse measurement signals. The resulting algorithm is termed as IAR-LASSO-ADMM-CFAR algorithm. Since the proposed IAR-LASSO-ADMM-CFAR algorithm could learn sparsity information at each iteration, it ensures a sparser and sparser solution and obtains the final sparse solution when the designed stopping criterion is satisfied. Finally, simulation results show that the proposed IAR-LASSO-ADMM-CFAR algorithm outperforms the conventional LASSO-ADMM algorithm and other existing algorithms in terms of reconstruction accuracy, and is more accurate than the existing algorithms in terms of sparsity order estimation.

This paper is organized as follows. Section II presents the signal model for CS, and describes the conventional LASSO-ADMM algorithm. The proposed IAR-LASSO-ADMM algorithm is described in detail in Section III. In Section IV, the effectiveness of the proposed IAR-LASSO-ADMM-CFAR algorithm is investigated by simulations. Finally, the conclusions are summarized in Section V.

## II. SYSTEM MODEL AND RELATED WORKS

In this section, the system model is firstly introduced, and then the LASSO-ADMM algorithm is described.

### A. System Model

During CS acquisition, an $N$-dimensional compressible signal $\mathbf{x}$ with sparsity order $k$ is directly sensed and compressed into an $M$-dimensional measurement vector $\mathbf{y}$ using an $M \times N$-dimensional sensing matrix $\mathbf{A}$. Mathematically, CS acquisition is represented as the following linear regression model [22]

$$\mathbf{y} = \mathbf{A}\mathbf{x} + \mathbf{n}, \quad (1)$$

where $\mathbf{n}$ is the measurement noise with zero mean and covariance matrix $\sigma^2 \mathbf{I}_M$. Here, $k < M \ll N$, $M \geq ck\log(N/k)$ for some constant $c$ and the compression ratio achieved is $N/M$. Let $I_{true} = \{n \in \{1, 2, \cdots, N\} : x_n \neq 0\}$ denote the nonzero support set of $\mathbf{x}$. Then, the sparsity level of $\mathbf{x}$ is $k = |I_{true}|$, where $|\cdot|$ denotes the cardinality of a set. An estimator produces an estimate $\hat{I}_{est} = \hat{I}(\mathbf{y})$ of $I_{true}$ based on the observed noisy vector $\mathbf{y}$ in (1). Given an estimator $\hat{I}_{est} = \hat{I}(\mathbf{y})$, its probability of error $p_{err} = \Pr(\hat{I}_{est} \neq I_{true})$ is taken with respect to randomness in $\mathbf{A}$, noise vector $\mathbf{n}$, and signal $\mathbf{x}$. The LASSO estimate of $\mathbf{x}$ is obtained by solving the following convex optimization [21]-[22]

$$\hat{\mathbf{x}} = \arg\min_{\mathbf{x}} \left(\frac{1}{2} \|\mathbf{y} - \mathbf{A}\mathbf{x}\|_2^2 + \lambda \|\mathbf{x}\|_1\right), \quad (2)$$

where $\lambda$ is a scalar nonnegative regularization parameter that regulates the nuclear norm of the solution $\hat{\mathbf{x}}$, hence leading to a sparser solution as $\lambda$ increases. Then, the nonzero components of $\hat{\mathbf{x}}$ can be utilized as an estimate of $I_{true}$.

### B. LASSO-ADMM Algorithm

In ADMM form, the LASSO problem in (2) can be written as [22]

$$\min f(\mathbf{x}) + g(\mathbf{z})$$
$$\text{Subject to } \mathbf{x} - \mathbf{z} = 0, \quad (3)$$

where $f(\mathbf{x}) = (1/2)\|\mathbf{A}\mathbf{x} - \mathbf{b}\|_2^2$ and $g(\mathbf{z}) = \lambda \|\mathbf{z}\|_1$. According to [22], the LASSO problem via ADMM (termed as LASSO-ADMM algorithm) is summarized in Algorithm 1 below.

---

**Algorithm 1:** LASSO-ADMM Algorithm

**Input:** Sampling vector $\mathbf{y} \in \mathrm{R}^M$, sensing matrix $\mathbf{A} \in \mathrm{R}^{M \times N}$, regularization parameter $\lambda$, augmented Lagrangian parameter $\rho$, relaxation parameter $\alpha$, absolute tolerance $\varepsilon^{abs}$, relative tolerance $\varepsilon^{rel}$, maximum iteration $t_{max}$;

---

**Output:** The sparsest solution $\hat{\mathbf{x}}$;

1: Initialization: $\mathbf{x}^1 = \mathbf{0}$, $\mathbf{z}^1 = \mathbf{0}$, $\mathbf{u}^1 = \mathbf{0}$;
2: for $k = 1 : t_{max}$
3:   1) Update $\mathbf{x}$
4:     $\mathbf{x}' = (\mathbf{A}^T\mathbf{A} + \rho\mathbf{I})^{-1}(\mathbf{A}^T\mathbf{y} + \rho(\mathbf{z}^k - \mathbf{u}^k))$; (4)
5:     $\mathbf{x}^{k+1} = \alpha\mathbf{x}' + (1-\alpha)\mathbf{z}^k$; (5)
6:   2) Update $\mathbf{z}$
7:     $\mathbf{z}^{k+1} = S_{\lambda/\rho}(\mathbf{x}^{k+1} + \mathbf{u}^k)$; (6)
8:   3) Update $\mathbf{u}$
9:     $\mathbf{u}^{k+1} = \mathbf{u}^k + \mathbf{x}^{k+1} - \mathbf{z}^{k+1}$; (7)
10:   4) Calculate the norm of the primal and dual residuals as follows:
11:     $r_{pri}^k = \|\mathbf{x}^{k+1} - \mathbf{z}^{k+1}\|_2$,
    $r_{dul}^k = \|-\rho(\mathbf{z}^{k+1} - \mathbf{z}^k)\|_2$; (8)
12:   5) Calculate primal and dual tolerance as follows:
    $\varepsilon_{pri}^k = \sqrt{n}\varepsilon^{abs} + \varepsilon^{rel}\max\{\|\mathbf{x}^{k+1}\|_2, \|\mathbf{z}^{k+1}\|_2\}$; (9)
    $\varepsilon_{dul}^k = \sqrt{n}\varepsilon^{abs} + \varepsilon^{rel}\|\rho\mathbf{u}^{k+1}\|_2$; (10)
13:   6) Stopping determination
14:     If ($r_{pri}^k < \varepsilon_{pri}^k$ and $r_{dul}^k < \varepsilon_{dul}^k$); (11)
15:     break;
16:   end
17: end
18: $\hat{\mathbf{x}} = \mathbf{x}^{k+1}$. (12)

---

In the above ADMM-LASSO algorithm, $\rho$ is the augmented Lagrangian parameter, $\alpha \in (0, 2)$ in (5) is a relaxation parameter [22]. The technique in (5) is called over-relaxation when $\alpha > 1$, and called under-relaxation when $\alpha < 1$. Conventionally, the regularization parameter $\lambda$ is set as $\lambda = 0.1\lambda_{max}$, where $\lambda_{max} = \|\mathbf{A}^T\mathbf{y}\|_\infty$. It should be noted that the ADMM-LASSO codes are available from the following Stanford University websites: https://web.stanford.edu/~boyd/papers/admm/lasso/lasso.html. However, the value of the regularization parameter $\lambda$ depends on the noise power. Therefore, the ADMM-LASSO algorithm usually provides poor results when the regularization parameter $\lambda$ remains fixed. In order to solve this problem, in next section we will propose a method to iteratively update the regularization parameter in the LASSO-ADMM algorithm based on the zero support of the estimates at each iteration.

### III. EFFICIENT IAR-LASSO-ADMM-CFAR ALGORITHM

In this section, we firstly propose a method to iteratively update the regularization parameter in the LASSO-ADMM algorithm, and provide a method to calculate the regularization parameter based on zero support of the estimates at each iteration. Then, we design a suitable stopping criterion to terminate the iteration process. Finally, the IAR-LASSO-ADMM-CFAR algorithm is described.

*A. Adaptive Update of the Regularization Parameter*

As discussed before, the regularization parameter $\lambda$ in the LASSO-ADMM algorithm depends on the noise power, and thus the choice of $\lambda$ greatly influences the behavior of the signal reconstruction from the noisy measurements. Therefore, it is required to find a suitable sequence of $\lambda$ which can be adapt to the varying sparsity order and different SNR. In this work, we propose an adaptive method to iteratively optimize the regularization parameter $\lambda$ by adding an outer iteration to the LASSO-ADMM algorithm. Specifically, at each iteration the complement of the nonzero support of the LASSO estimate $\hat{\mathbf{x}}$ in (12) is utilized to estimate the noise variance, and the estimated noise variance is employed to formulate an estimate for $\lambda$, which will be described in the following.

At the $l$th outer iteration, it is assumed that the LASSO estimate $\hat{\mathbf{x}}$ in (12) of $\mathbf{x}$ is denoted as $\hat{\mathbf{x}}^l$, $l = 0, 1, 2, \cdots$. Let $\hat{I}_{est}^l = \{n \in \{1, 2, \cdots, N\} : \hat{x}^l \neq 0\}$ denote the nonzero support set of $\hat{\mathbf{x}}^l$, and its complement is denoted as $\hat{I}_{est}^{l,c} = \{n \in \{1, 2, \cdots, N\} : \hat{x}^l = 0\}$. In addition, $|\hat{I}_{est}^l|$ and $|\hat{I}_{est}^{l,c}|$ are the cardinality of $\hat{I}_{est}^l$ and $\hat{I}_{est}^{l,c}$, respectively. Initially, the initial value $\lambda^0$ for the regularization parameter $\lambda$ should be set sufficiently small such as $\lambda^0 = 0.1\lambda_{\max}$ to guarantee that $\hat{I}_{est}^{l,c}$ only corresponding to the noise terms (i.e., zero support set) of $\hat{\mathbf{x}}^l$, where $\lambda_{\max} = \|\mathbf{A}^T\mathbf{y}\|_\infty$. From (1), we calculate

$$\tilde{\mathbf{x}} = \mathbf{A}'\mathbf{y}. \tag{13}$$

Then, we define $\tilde{\mathbf{x}}_{\hat{I}_{est}^{l,c}}$ as the subset corresponding to the zero support of $\tilde{\mathbf{x}}$ as follows:

$$\tilde{\mathbf{x}}_{\hat{I}_{est}^{l,c}}(k) = \begin{cases} \tilde{x}(k), & k \in \hat{I}_{est}^{l,c} \\ 0, & k \in \hat{I}_{est}^l \end{cases}, \quad k = 0, 1, \cdots, N, \tag{14}$$

where $\tilde{x}(k)$ is the $k$th components of $\tilde{\mathbf{x}}$ in (13). Then, we calculate

$$\tilde{\mathbf{y}} = \mathbf{A}\tilde{\mathbf{x}}_{\hat{I}_{est}^{l,c}}. \tag{15}$$

As the sensing matrix $\mathbf{A}'$ is orthonormal and $\tilde{\mathbf{x}}_{\hat{I}_{est}^{l,c}}$ contains only noise terms, it can be easily inferred that the elements of $\tilde{\mathbf{y}}$ can be regarded as normal random variables. Therefore, the noise variance $\sigma^2$ can be estimated from $\tilde{\mathbf{y}}$ as follows:

$$\hat{\sigma}_l^2 = \frac{1}{2|\hat{I}_{est}^{l,c}|} \sum_{j=1}^{M} |\tilde{y}(j)|^2, \tag{16}$$

where $\tilde{y}(j)$ is the $j$th component of $\tilde{\mathbf{y}}$. Then, we define a subset $X$ corresponding to the nonzero indexes set $\hat{I}_{est}^l$ of $\hat{\mathbf{x}}^l$ as follows:

$$X = [|\tilde{\mathbf{x}}(\hat{I}_{est}^l(1))|, |\tilde{\mathbf{x}}(\hat{I}_{est}^l(2))|, \cdots, |\tilde{\mathbf{x}}(\hat{I}_{est}^l(\hat{I}_{est}^l))|]. \tag{17}$$

When assuming that no signals exist, it is indicated in [14] that $X$ can be regarded as a random variable sample set that follows a Rayleigh distribution with a probability distribution function (PDF) given by

$$f_X(z) = \frac{x}{\sigma^2} e^{-z^2/2\sigma^2} \tag{18}$$

and a cumulative distribution function (CDF) given by

$$F_X(z) = 1 - e^{-z^2/2\sigma^2} \tag{19}$$

where $\sigma$ is the scale parameter of $X$, and it can be estimated using (16). Using the estimated scale parameter $\hat{\sigma}_l^2$ in (16), the desired false alarm rate $P_{fa}$ is related to CDF as follows:

$$P_{fa} = 1 - F_X(T_{fa}) = e^{-T_{fa}^2/(2\hat{\sigma}_l^2)}. \tag{20}$$

According to (20), the false alarm regulation threshold $T_{fa}$ of the proposed algorithm is given by [14]

$$T_{fa} = \sqrt{-2\hat{\sigma}_l^2 \log P_{fa}}. \tag{21}$$

Using the $T_{fa}$, we update the $\hat{I}_{est}^l$ to $\hat{I}_{est}^{l(u)} = \{n \in \hat{I}_{est}^l : |\hat{\mathbf{x}}^l(n)| > T_{fa}\}$. That is to say, the updated index set $\hat{I}_{est}^{l(u)}$ is obtained by remaining the indexes in $\hat{I}_{est}^l$ that satisfies $|\hat{\mathbf{x}}^l(n)| > T_{fa}$, where $n \in \hat{I}_{est}^l$. Then, we update the estimates for $\hat{\mathbf{x}}^l$ as

$$\tilde{\mathbf{x}}^{l(u)}(k) = \begin{cases} \hat{x}^l(k), & k \in \hat{I}_{est}^{l(u)} \\ 0, & k \notin \hat{I}_{est}^{l(u)} \end{cases}, \quad k = 1, 2, \cdots, N. \tag{22}$$

Then, we update the initial value $\mathbf{x}^1$ of the inner LASSO-ADMM algorithm as $\mathbf{x}^1 = \tilde{\mathbf{x}}^{l(u)}$ and the regularization parameter $\lambda$ as $\lambda = T_{fa}$, and proceed to perform the next iteration of the LASSO-ADMM algorithm. Since this algorithm could learn sparsity information at each iteration, it ensures a sparser and sparser solution. In fact, a suitable stopping criterion is important for signal reconstruction, as stopping too early would increase false alarms, while stopping late would increase miss-detections. In next subsection, we will propose a stopping criterion.

*B. Design of stopping Criterion*

According to (16), at the $l$ th outer iteration, the noise variance is estimated as $\hat{\sigma}_l^2$. If $\hat{\sigma}_l^2 > \hat{\sigma}_{l-1}^2$, it can be inferred that noise components may still exist in the estimated $\tilde{\mathbf{x}}^{l(u)}$. Therefore, it should update the regularization parameter $\lambda$ and continue to perform the next inner iteration to further eliminate the noise components in the measurements. Otherwise, i.e., if $\hat{\sigma}_l^2 \leq \hat{\sigma}_{l-1}^2$, it can be inferred that the signal components are wrongly detected as the noise components. Therefore, in this case the outer iteration should terminate and the estimate $\tilde{\mathbf{x}}^{l-1(u)}$ in the previous iteration should be taken as the final estimate for $\mathbf{x}$.

Based on above discussion, the IAR-LASSO-ADMM-CFAR algorithm can be summarized in Algorithm 2 below. The proposed algorithm composed of an inner iteration (i.e., LASSO-ADMM algorithm) and an out iteration. Specifically, at each outer iteration the zero support of the estimate obtained by the inner LASSO-ADMM algorithm is utilized to estimate the noise variance, and the noise variance is utilized to update the threshold according to a pre-defined const false alarm rate (CFAR). Then, the resulting threshold is utilized to update both the non-zero support of the estimate and the regularization parameter, and proceed to the next inner iteration. In addition, a suitable stopping criterion is designed to terminate the outer iteration process to obtain the final non-zero support of the estimate of the sparse measurement signals. Since the proposed IAR-LASSO-ADMM-CFAR algorithm could learn sparsity information at each iteration, it ensures a sparser and sparser solution and obtains the final sparse solution when the designed stopping criterion is satisfied.

---

**Algorithm 2:** AIR-LASSO-ADMM-CFAR Algorithm

**Input:** Sampling vector $\mathbf{y} \in \mathrm{R}^M$, sensing matrix $\mathbf{A} \in \mathrm{R}^{M \times N}$, regularization parameter $\lambda$, augmented Lagrangian parameter $\rho$, relaxation parameter $\alpha$, absolute tolerance $\varepsilon^{\mathrm{abs}}$, relative tolerance $\varepsilon^{\mathrm{rel}}$, maximum iteration $t_{\max}$;

**Output:** The sparsest solution $\hat{\mathbf{x}}$;

1: Initialization: $\mathbf{x}^1 = \mathbf{0}$, $\mathbf{z}^1 = \mathbf{0}$, $\mathbf{u}^1 = \mathbf{0}$, $\lambda$ (e.g., $\lambda = 0.1 \| \mathbf{A}^T \mathbf{y} \|_\infty$);
2: Set the count of outer iterations $l = 0$
3: **Do**  // outer iteration
4: for $k = 1 : t_{\max}$  // inner iteration
5:   1) Update $\mathbf{x}$
6:     $\mathbf{x}' = (\mathbf{A}^T \mathbf{A} + \rho \mathbf{I})^{-1}(\mathbf{A}^T \mathbf{y} + \rho(\mathbf{z}^k - \mathbf{u}^k))$; (23)
7:     $\mathbf{x}^{k+1} = \alpha \mathbf{x}' + (1-\alpha) \mathbf{z}^k$; (24)
8:   2) Update $\mathbf{z}$
9:     $\mathbf{z}^{k+1} = S_{\lambda/\rho}(\mathbf{x}^{k+1} + \mathbf{u}^k)$; (25)
10:   3) Update $\mathbf{u}$
11:     $\mathbf{u}^{k+1} = \mathbf{u}^k + \mathbf{x}^{k+1} - \mathbf{z}^{k+1}$; (26)
12:   4) Calculate the norm of the primal and dual residuals as follows:
13:     $r_{\mathrm{pri}}^k = \| \mathbf{x}^{k+1} - \mathbf{z}^{k+1} \|_2$, (27)
    $r_{\mathrm{dul}}^k = \| -\rho(\mathbf{z}^{k+1} - \mathbf{z}^k) \|_2$; (28)
14:   5) Calculate primal and dual tolerance as follows:
    $\varepsilon_{\mathrm{pri}}^k = \sqrt{n}\varepsilon^{\mathrm{abs}} + \varepsilon^{\mathrm{rel}} \max\{\| \mathbf{x}^{k+1} \|_2, \| \mathbf{z}^{k+1} \|_2\}$; (29)
    $\varepsilon_{\mathrm{dul}}^k = \sqrt{n}\varepsilon^{\mathrm{abs}} + \varepsilon^{\mathrm{rel}} \| \rho \mathbf{u}^{k+1} \|_2$; (30)
15:   6) Stopping determination
16:     If ($r_{\mathrm{pri}}^k < \varepsilon_{\mathrm{pri}}^k$ and $r_{\mathrm{dul}}^k < \varepsilon_{\mathrm{dul}}^k$); (31)
17:     break;
18:   end
19: end
20:   $\hat{\mathbf{x}}_{\mathrm{inner}} = \mathbf{x}^{k+1}$; (32)
21:   $\hat{\mathbf{x}}^l \leftarrow \hat{\mathbf{x}}_{\mathrm{inner}}$; (33)
22: 7) Calculate the nonzero and zero support set of $\hat{\mathbf{x}}^l$
    $\hat{I}_{est}^l = \{n \in \{1,2,\cdots,N\} : \hat{\mathbf{x}}^l \neq 0\}$; (34)
    $\hat{I}_{est}^{l,c} = \{n \in \{1,2,\cdots,N\} : \hat{\mathbf{x}}^l = 0\}$; (35)
23: 8) Calculate the residual noise in the measurement $\mathbf{y}$
  a) Calculate $\tilde{\mathbf{x}} = \mathbf{A}'\mathbf{y}$
  b) Calculate the noise components $\tilde{\mathbf{x}}_{\hat{I}_{est}^{l,c}}$ by using (14);
  c) Calculate the residual noise $\tilde{\mathbf{y}} = \mathbf{A}\tilde{\mathbf{x}}_{\hat{I}_{est}^{l,c}}$ by using (15);
24: 9) Estimate the noise variance $\hat{\sigma}_l^2$ by using (16);
25: 10) Calculate the regulation threshold $T_{fa}$ by using (21);
26: 11) Update the nonzero support set:
    $\hat{I}_{est}^{l(u)} = \{n \in \hat{I}_{est}^l : |\hat{\mathbf{x}}^l(n)| > T_{fa}\}$; (36)
27: 12) Update the estimates $\hat{\mathbf{x}}^l$ as $\tilde{\mathbf{x}}^{l(u)}(k)$ by using (22);
28: 13) If $\hat{\sigma}_l^2 > \hat{\sigma}_{l-1}^2$ then
  a) Update the regulation parameter $\lambda \leftarrow T_{fa}$;
  b) Go to 4, and proceed to the next inner iteration;
else
  a) Obtain the final estimate $\tilde{\mathbf{x}}^{l(u)}(k)$, and go to 30;
end
29: **end**

30: Output: $\hat{\mathbf{x}} = \tilde{\mathbf{x}}^{l(u)}(k)$. (37)

## IV. SIMULATION RESULTS AND DISCUSSIONS

In this section, simulations are presented to illustrate the performance of the proposed algorithm, and compare the performance of the proposed algorithm with that of other existing algorithms for problems (1). In our simulations, in addition to the proposed algorithm, we consider the following algorithms: l1-ls algorithm in [33], monotone GPSR algorithm in [19], and LASSO-ADMM algorithm and LASSO-LSQR algorithm in [22]. The MATLAB implementations of the above algorithms are freely available at http://www.lx.it.pt/~mtf/GPSR/ for the GPSR algorithm, at http://www.stanford.edu/~boyd/l1_ls for the l1-ls algorithm, and at https://web.stanford.edu/~boyd/papers/admm/lasso/lasso.html for the LASSO-ADMM algorithm and LASSO-LSQR algorithm. All the experiments were carried out on a personal computer with an Intel(R) Core (TM) i5-8250U 1.6 GHz processor and 16GB of memory, using a MATLAB implementation of the proposed algorithm. The parameters of the proposed algorithm and the LASSO-ADMM algorithm are set as follows: the augmented Lagrangian parameter $\rho = 0.9$, the relaxation parameter $\alpha = 1.5$, the absolute tolerance $\varepsilon^{abs} = 10^{-5}$, the relative tolerance $\varepsilon^{rel} = 10^{-4}$, and the maximum iteration $t_{max} = 1000$. For the regularization parameter $\lambda$, its initial value is set as $\lambda = 0.1\lambda_{max}$, where $\lambda_{max} = \|\mathbf{A}^T\mathbf{y}\|_\infty$. In addition, the variables $\mathbf{u}^0$ and $\mathbf{z}^0$ are initialized to be zero.

In order to quantify the reconstruction accuracy of the proposed algorithm, the conventional mean square error (MSE) is calculated as:

$$\text{MSE} = \|\hat{\mathbf{x}} - \mathbf{x}\|_2^2 / N, \quad (38)$$

where $\mathbf{x}$ is the original signal and $\hat{\mathbf{x}}$ is the reconstructed signal. In all simulations, the signal-to-noise ratio (SNR) will be defined as

$$\text{SNR} = 1/\sigma^2. \quad (39)$$

### A. Comparisons of estimation performance of various algorithms when the sparsity order is fixed

In this simulation, we compare the estimation performance of the proposed algorithm against the existing algorithms when the sparsity order is fixed. Let $\mathbf{A}$ be a random $M \times N$ sensing matrix, with $M = 1024$ and $N = 4096$. In simulations, we choose $\mathbf{y} = \mathbf{A}\mathbf{x}_{true} + \mathbf{n}$, where $\mathbf{x}_{true}$ is a vector with $k = 150$ randomly placed $\pm 1$ spikes and with zeros in the other components, and $\mathbf{n}$ is a Gaussian white noise vector with variance $\sigma^2$. The regularization parameter of the l1-ls algorithm in [33] is set as $\lambda_{l1-ls} = 0.2\lambda_{max}$, and that of the GPSR algorithm in [19] and LASSO-ADMM algorithm and LASSO-LSQR algorithm in [22] is set as $\lambda = 0.1\lambda_{max}$, the standard deviation of the noise is set as $\sigma = 0.05$, and the false alarm rate is set as $P_{fa} = 0.001$. In addition, 50 random data sets, i.e., triplets $(\mathbf{x}, \mathbf{y}, \mathbf{A})$, are generated for sparse estimation. Then, the sparsity order estimate, MSE (with respect to the true $\mathbf{x}_{true}$), the value of objective function, and the CPU times are obtained by using various algorithms. Finally, the results for the MSE, the CPU times, the value of objective function and the sparsity order estimate are obtained by averaging over 50 independent runs.

Table I reports the sparsity order estimate, CPU times, the MSE, and the value of the objective function of various algorithms (average over 50 independent runs) when the true sparsity order $k = 150$, $P_{fa} = 0.001$ and noise variance $\sigma = 0.05$. As can be seen from these results, though the proposed IAR-LASSO-ADMM-CFAR algorithm is slightly inferior to the l1-ls, GPSR-basic, LASSO-ADMM and LASSO-LSQR in terms of CPU time, MSE and the value of objective function, it is far better than these algorithm in terms of sparsity order estimation. That is to say, the sparsity order estimate $\hat{k}$ of the proposed algorithm is closest to the true value $k = 150$, while the sparsity order estimate $\hat{k}$ of other algorithms is far away from the true value $k = 150$.

TABLE I
The sparsity order estimate, CPU times, the MSE, and the value of objective function of various algorithms (average over 50 independent runs) when the true sparsity order $k = 150$, $P_{fa} = 0.001$ and noise variance $\sigma = 0.05$.

| Algorithm | CPU time (seconds) | MSE | value of objective function | sparsity order estimate |
|---|---|---|---|---|
| l1-ls | 1.5797 | 0.0049 | 14.5079 | 4096 |
| GPSR-basic | 0.1619 | 0.0062 | 7.2493 | 480.16 |
| LASSO-ADMM | 2.3200 | 0.0047 | 7.2313 | 431.72 |
| LASSO-LSQR | 5.0597 | 0.0052 | 7.2347 | 448.98 |
| IAR-LASSO-ADMM-CFAR | 9.3666 | 0.0139 | 14.0291 | 151.58 |

### B. Comparisons of estimation performance of various algorithms under various SNR

In this simulation, we compare the estimation performance of various algorithms under various SNR. Here, the SNR is defined as $\text{SNR} = 1/\sigma^2$. We consider similar settings as the one in the first experiment above, but with only one difference. Instead of being set as a fixed value, the standard deviation of the noise $\sigma$ is set as a varying value. In addition, 50 random data sets, i.e., triplets $(\mathbf{x}, \mathbf{y}, \mathbf{A})$, are generated, and the results are obtained by averaging over 50 independent runs.

Fig. 1 (a) and Fig. 1(b), respectively, shows the average MSE and average sparsity order estimate of various algorithms (average over 50 independent runs) as a function of SNR. As can be seen from Fig. 1 (a), the MSE of the proposed IAR-LASSO-ADMM-CFAR algorithm is far less than that of the other algorithms when the SNR is smaller than 18dB, while will become slightly larger than that of the other algorithms when the SNR is greater than 18dB. As can be seen from Fig. 2(b), the sparsity order estimate $\hat{k}$ of the proposed algorithm is closest to the true value $k = 150$ among these algorithms. Moreover, the sparsity order estimate $\hat{k}$ of the proposed algorithm is almost the same as the true value $k = 150$ when the SNR is greater than 25dB (i.e., $\sigma \leq 0.056$).

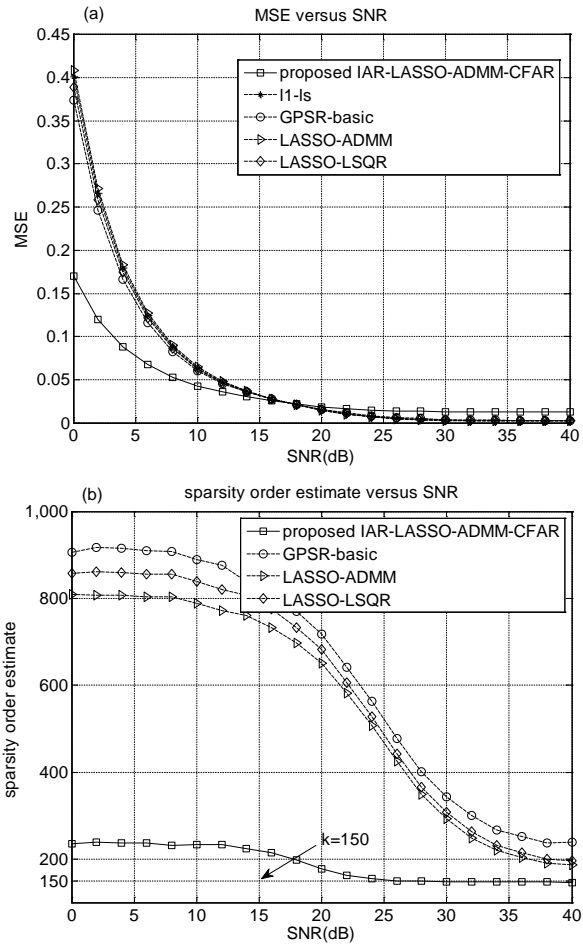

Fig. 1. (a) Average MSE and (b) average sparsity order estimate of various algorithms (average over 50 independent runs) as a function of SNR.

## C. Comparisons of estimation performance of various algorithms under various sparsity order

In this simulation, we compare the estimation performance of various algorithms under various sparsity order. We consider similar settings as the one in the first experiment above, but with the following differences. The standard deviation of the noise is set as $\sigma = 0.01$, and the false alarm rate is set as $P_{fa} = 0.001$. Instead of setting a fixed sparsity order, we consider a range of degrees of sparsity: the number $k$ of nonzero spikes in $\mathbf{x}$ (randomly located values of $\pm 1$) ranges from 4 to 240. For each value of $k$, 50 random data sets, i.e., triplets $(\mathbf{x}, \mathbf{y}, \mathbf{A})$, are generated. Then, the sparsity order estimate and the MSE (with respect to the true $\mathbf{x}$) are obtained by using various algorithms. Finally, the results for the MSE and the sparsity order estimate are averaged over 50 independent runs.

Fig. 2 (a) and Fig. 2(b), respectively, shows the average MSE and average sparsity order estimate of various algorithms (average over 50 independent runs) as a function of the true sparsity order. As can be seen from Fig. 2, though Fig. 2(a) shows that the proposed IAR-LASSO-ADMM-CFAR algorithm has larger average MSE than the l1-ls, the GPSR-basic, the LASSO-ADMM and the LASSO-LSQR, Fig. 2(b) shows that the sparsity order estimate of the proposed IAR-LASSO-ADMM-CFAR algorithm is closest to the true value $k$ among these algorithms. Moreover, Fig. 2(b) shows that the sparsity order estimate of the proposed IAR-LASSO-ADMM-CFAR algorithm is nearly the same as the true value $k$ when the true value is less than 150, and is slightly less than the true value when the true value is greater than 150.

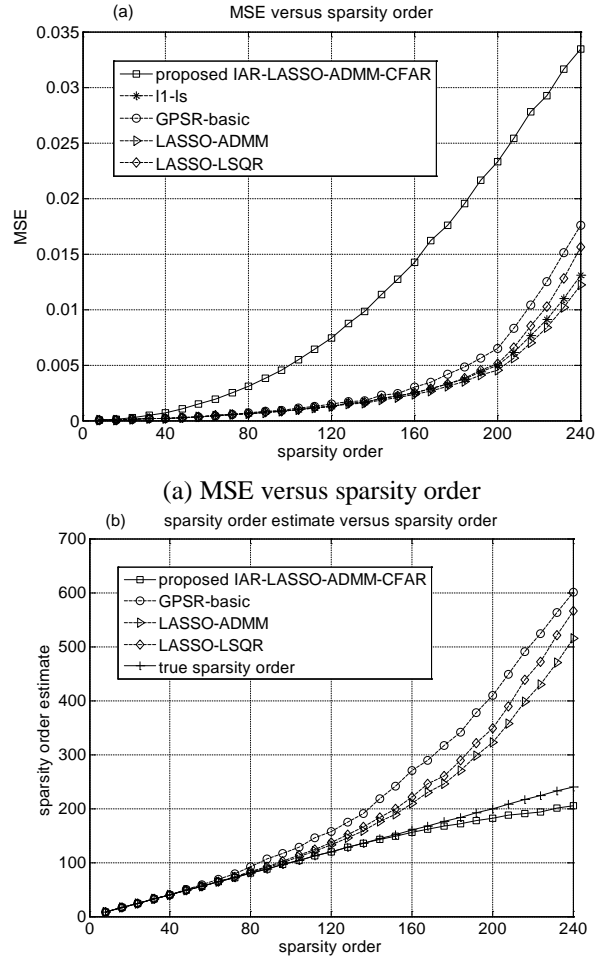

(a) MSE versus sparsity order

(b) Sparsity order estimate versus the true sparsity order

Fig. 2. (a) average MSE and (b) average sparsity order estimate of various algorithms (average over 50 independent runs) as a function of the true sparsity order.

ACKNOWLEDGMENT

This work was supported by Shanghai Fundamental Research Key Project under Grant 20JC1416504.